# Diamond open access and open infrastructures have shaped the Canadian scholarly journal landscape since the start of the digital era

Simon van Bellen[1], Lucía Céspedes[1,2]

[1] Consortium Érudit, Montréal, QC, Canada
[2] UNESCO Chair on Open Science, Université de Montréal, Montréal, QC, Canada

## Abstract

Scholarly publishing involves multiple stakeholders having various types of interest. In Canada, the implication of universities, the presence of societies and the availability of governmental support for periodicals seem to have contributed to a rather diverse ecosystem of journals. This study presents in detail the current state of these journals, in addition to past trends and transformations during the 20th century and, in particular, the digital era. To this effect, we created a new dataset, including a total of 1256 journals, 944 of which appeared to be active today, specifically focusing on the supporting organizations behind the journals, the types of (open) access, disciplines, geographic origins, languages of publication and hosting platforms and tools. The main overarching traits across Canadian scholarly journals are an important presence of Diamond open access, which has been adopted by 62% of the journals, a predominance of the Social Sciences and Humanities disciplines and a scarce presence of the major commercial publishers. The digital era allowed for the development of open infrastructures, which contributed to the creation of a new generation of journals that massively adopted Diamond open access, often supported by university libraries. However, journal cessation also increased, especially among the recently founded journals. These results provide valuable insights for the design of tailored practices and policies that cater to the needs of different types of periodicals and that take into account the evolving practices across the Canadian scholarly journal landscape.

## Keywords

# Introduction

In today's increasingly internationalized scientific field, it may well seem that Louis Pasteur's adage about science knowing no country is truer than ever. However, to paraphrase the pioneering French biochemist, scholarly journals certainly do. At least, a specific group within scholarly periodical publications: that of national journals. These can be defined as journals where researchers from a particular country publish predominantly in order to communicate their research to peers and other interested audiences from that country; and where the actors involved in publication, such as the editorial team and supporting organisms, share a common geographical location (Lange & Severson, 2021; Moed et al., 2021). National, or domestic journals are also likely to present research topics with a geographically limited scope or having a specific, national or local context (Fortin, 2018; Gingras & Mosbah-Natanson, 2010; Ma, 2019). Thereby, they may be considered opposing "international" journals, which may attract international authorships and readerships, present universal research topics, and often be associated with the Natural and Health Sciences (Gingras, 2014) and multinational, commercial publishers (Larivière et al., 2015).

National journals are the natural channels for the creation of regionally relevant research communities and scholarly networks, especially in the Social Sciences and Humanities (SSH) (Sivertsen, 2016). Because domestic journals often publish in a national language, other than English, they contribute to the overall multilingualism of the scientific field and act as a counterweight to the hegemony of English as the *de facto* sole international language of science (Céspedes, 2021; Pölönen et al., 2021). Language is also a facilitator to attract a wider readership beyond academia, a relevant issue since the contents of national journals often cover research agendas of particular interest for local stakeholders such as social organizations, policymakers, and funders.

As the term "national" or "local" has come to often connote lesser value as opposed to "international" (Lillis, 2012), national journals are often disregarded by major publishers - and some researchers alike. Canadian scholars, especially in the SSH, tend to shift their research scope, aiming for a less specific and more international focus (Larivière & Warren, 2019), which is likely to affect the Canadian scholarly journal landscape. Previous studies have endeavoured to characterize it, ever since the pre-digital era (Gordon, 1984) or during its early years (Boismenu & Beaudry, 2002). In the last decade, the Canadian scholarly journal landscape has been described as highly heterogeneous, with "[...] small and some mid-sized journals that publish quality research findings and other scholarship on a diverse range of subjects" (Paquin, 2016), characterized by a high prevalence of different modalities of open access (OA) (Larivière et al., 2021).

The variety of agents involved in supporting journals is a distinguishing trait of the Canadian scholarly publishing ecosystem. Unlike the United States and Great Britain, Canada does not have a dominance of domestic commercial publishers and internationally oriented journals (Boismenu & Beaudry, 2002; Moed et al., 2021); instead, the majority of Canadian journals are supported by research institutions, professional and learned societies, university libraries and presses, public funders, and volunteers (Lange & Severson, 2021; Larivière et al., 2021; Paquin, 2016). The Social Sciences and Humanities Research Council (SSHRC) has provided financial support to Canadian scholarly journals in these research areas since 1979, while the ancestors of the current *Fonds de Recherche du Québec* (FRQ) have supported Quebec SSH journals since, at least, 1981[1]. Furthermore, journals hosted by Érudit, a Quebec-based platform for the dissemination of, mostly, SSH journals, benefit from funding through its Partnership for Open Access, which, as of 2024, redistributes the contributions of more than 90 libraries to support over 250 non-profit journals. There is a consensus that this support is crucial to ensure the sustainability of non-profit journals, particularly in face of the competition of commercial publishers (Canadian Scholarly Publishing Working Group, 2017), which offer perceived benefits related to infrastructure, indexing, marketing and editorial training opportunities, in addition to prestige (Fyfe et al., 2017; Krapež, 2023).

Non-profit journals are considered fragile for a number of reasons. Characterized by low publishing volumes, the support they receive is not always stable, and these journals rely frequently on volunteers and unpaid labour for their everyday activities (Björk et al., 2016; Bosman et al., 2021; Lange & Severson, 2022; Morrison, 2016). Besides, OA mandates may have the undesired side effect of exacerbating the difficulties in their financing. As mentioned, commercial publishers, even if they have a lower presence among Canadian journals, are competitors for many national scholar-led journals (Lange & Severson, 2021). French-language publishing has to compete with English, which is not only the 'other' national language, but also the 'international language of science'. Still in Canada, the growth of English in scholarly publishing has been affecting French-language publishing in most fields for multiple decades, except in the Arts and Humanities, for which the decline has been more recent (Larivière, 2018). Finally, national journals are often underrepresented in the traditional bibliometric databases, such as Web of Science and Scopus, both at journal and article level, quantitatively as well as qualitatively (Basson et al., 2022; Larivière, 2018). As a result, the value of these journals tends to be underestimated, or is simply not accounted for, in research evaluation.

Due to the highly dynamic nature of scholarly publishing, the total volume of Canadian journals is difficult to establish and estimations may vary according to the sources; however, several efforts have been made. In the early 2000s, in a context of transition to online publishing and gradual adoption of new digital technologies, Lorimer and Lindsay (2004) estimated around 150-200 active Humanities and Social Sciences journals in Canada. Years later, Paquin (2016) surveyed 337 Canadian journals, concluding 25% of journals had adopted OA, with modest annual revenues, ranging between 30,000 and 80,000 CA $. More recently, Larivière et al. (2021) collected data from Ulrich's Periodicals Directory and retained 825 active Canadian journals in 2019, of which almost three-quarters published in the Humanities and Social Sciences (611 journals) and around a quarter in Science, Technology and Medicine (214 journals). Focusing on independent journals, defined as those not affiliated with commercial publishers, Lange and Severson (2021) identified 485 Canadian scholarly publications across all subject areas.

---

[1] We have searched online to recover past versions of the program. The earliest traces of this program dated back to 1981: Rapport annuel 1981-1982 / FCAC, Fonds F.C.A.C. pour l'aide et le soutien à la recherche. Downloaded from https://numerique.banq.qc.ca/patrimoine/details/52327/4274170.

Considering the various models for journal support, supporting actors and the rapidly evolving publishing practices and policies, an up-to-date characterization of the Canadian scholarly journal landscape may serve as a valuable tool to establish strategies for optimizing journal support, thus assisting universities, funders and infrastructures. It may also help journal editors in understanding their environment and identifying their journals' "niche".

Thus, the aim of this study is to characterize the Canadian scholarly periodical publishing landscape, based on an up-to-date list of Canadian scholarly journals. To this effect, we did not *a priori* define journal types, which means that various, sometimes overlapping, types of journals were included in the analyses, such as national, non-profit, commercial and student journals. This approach allowed us to describe and distinguish the various journal types actively being published from within Canada. Generally, we focused on the organizations behind the journals, the types of OA, disciplines, geographic origins, languages and dissemination platforms and tools. In contrast with previously mentioned studies, the analyses integrated a historical perspective, by the identification of the journals' years of founding and cessation, which allowed for the detection of tendencies in the evolution of journal characteristics.

# Methods

Journal titles and their characteristics were compiled using a variety of sources. Four criteria were established for journals in order to be included in the analysis: 1) the journal should be scholarly, including peer- or editorial review, and periodical, thus excluding proceedings; 2) the journal should be mainly managed from within a Canada-based institution, association or society; 3) the journal should have an ISSN associated; 4) the journal should appear legitimate, i.e. journals should not be associated with publishers known to have 'questionable' practices. Student journals were also included if they responded to these criteria. The main input for journal titles and ISSN were the dataset created by Larivière et al. (2021), OpenAlex' Sources dataset (Priem et al., 2022), the open dataset published by Lange and Severson and updated in 2021 (Lange & Severson, 2019), CRKN's Open Access Journals List[2], as well as internal data from the Public Knowledge Project and Érudit. Mir@bel (https://reseau-mirabel.info/) was used occasionally to add lacking information. After combining these datasets, duplicates were removed based on ISSN, with additional checking on the journal titles. The dataset is available on Érudit's Borealis Dataverse (Van Bellen, 2024), both the snapshot used for analyses and a version being updated regularly.

Both active and ceased journals were included in the study. Journals were classified as ceased in case no volume had been published since 2020. In the process, several journals were found to be part of a sequence. For example, the journal of the Mineralogical Association of Canada, today titled *Canadian journal of mineralogy and petrology*, was published as *Canadian Mineralogist* between 1957 and 2022, with both titles bearing different ISSNs. As both the organization managing and the scope of the journal remained unchanged, we considered these cases as representing a single periodical. This allowed for a more coherent analysis of historical trends in publishing.

---

[2] Available at https://www.crkn-rcdr.ca/en/crkn-open-access-journals-list

Most fields of the dataset required additional resources to complete lacking data or to validate. Library and Archives Canada's Aurora catalog was used to obtain the years of creation and cessation (if applicable) of the journals and to validate the organization managing the journal. It was also valuable in obtaining information on journal sequences. The journals' accepted languages for submissions were documented based on existing datasets mentioned earlier, or they were identified on their respective websites. OpenAlex was used as the primary source for OA status, however, we validated output manually as some information appeared inaccurate. Each journal was classified as Gold OA, Diamond OA, Hybrid, or Subscription, following the definitions by Piwowar et al. (2018)[3]. Finally, journal websites were consulted to verify access types and to obtain journal policies relative to languages accepted. Internet Archive's Wayback Machine (https://web.archive.org/) proved helpful in retrieving language policies and other details of ceased journals. Each journal was manually classified as belonging to one of nine fields: Arts and Literature, Economics and Management, Health Sciences, Humanities, Natural Sciences, Professional Fields, Psychology, Social Sciences and Pluridisciplinary. These fields have been used by Larivière et al. (2021) and reflect the distribution of disciplines most common for Canadian journals. The Pluridisciplinary field was used for journals that accept submissions of any discipline.

## Results

### General overview

A total of 1256 Canadian scholarly journals were identified, of which 944 appeared to be actively publishing today. All provinces have a presence of active journals, but no journals were found to originate from the Northwest Territories or Nunavut (Table 1). There is a strong concentration of Canadian journal publishing in Ontario and Quebec, with 501 and 179 active journals, respectively. This speaks of a pattern of scientific, academic and editorial centers and peripheries within the country. Ontario, as the most populous province and seat of the federal government, is home to many learned societies and professional associations operating at the national level, in addition to the presence of older, established universities and research centers, well endowed with human and financial resources. The high publishing activity in Quebec is likely associated with the importance of French, and may have been further enhanced by FRQSC's journal support program; eligible journals must publish at least 50% of their articles in French.

---

[3] For a discussion on the working definitions of OA, particularly, the Diamond OA category, see Simard et al. (2024).

Table 1: Journal origin according to province or territory.

| Province/territory | Active journals | |
|---|---|---|
| | n | % |
| Ontario | 501 | 53 |
| Quebec | 179 | 19 |
| Alberta | 94 | 10 |
| British Columbia | 82 | 9 |
| Nova Scotia | 23 | 2 |
| Manitoba | 22 | 2 |
| Newfoundland and Labrador | 15 | 2 |
| New Brunswick | 14 | 1 |
| Saskatchewan | 11 | 1 |
| Prince Edward Island | 2 | <1 |
| Yukon | 1 | <1 |
| Total | 944 | 100 |

The current journal landscape is dominated by the Social Sciences and Humanities and connected disciplines, which together account for 71% of the journals. Health and Natural Sciences represent 15% and 10% of the journals, respectively, while 4% of the journals were classified as Pluridisciplinary.

Table 2: Active journals according to disciplinary field.

| Discipline | Active journals - Total | | Active journals - Six major publishers | |
|---|---|---|---|---|
| | n | % | n | % |
| Social Sciences | 220 | 23 | 10 | 18 |
| Professional Fields | 173 | 18 | 5 | 9 |
| Health Sciences | 141 | 15 | 15 | 27 |
| Humanities | 126 | 13 | 3 | 5 |
| Natural Sciences | 98 | 10 | 13 | 23 |
| Arts and Literature | 92 | 10 | 0 | 0 |
| Pluridisciplinary | 38 | 4 | 0 | 0 |
| Economics and Management | 30 | 3 | 8 | 14 |
| Psychology | 26 | 3 | 2 | 4 |
| Total | 944 | 100 | 56 | 100 |

The six major commercial publishers globally, which include RELX-Elsevier, Springer Nature, Wiley, MDPI, Taylor & Francis and Sage Publishing (Van Bellen et al., 2024), together account for 56 journals, or close to 6% of the active journals. Compared to other journals, these journals are often associated with the Natural Sciences, the Health Sciences and Economics and Management (Table 2).

Currently active scholarly periodicals are characterized by a high diversity in access types, infrastructures for management and dissemination and hosting organizations. Since many variables showed specific interrelationships, we used multivariate analyses to identify the major patterns ('gradients') present within the journal ecosystem. The main gradient opposes library-supported Diamond OA journals, which use OJS, and are associated with the Social Sciences, on one hand, and Gold OA and hybrid journals of the Natural and Health Sciences, typically published by learned societies and major commercial publishers, on the other (Figure 1). This main gradient also showed strong journal divergence according to the year of creation of the journal. Multivariate analysis was further used to create a typology of four groups of journals, based on hierarchical clustering (Table 3).

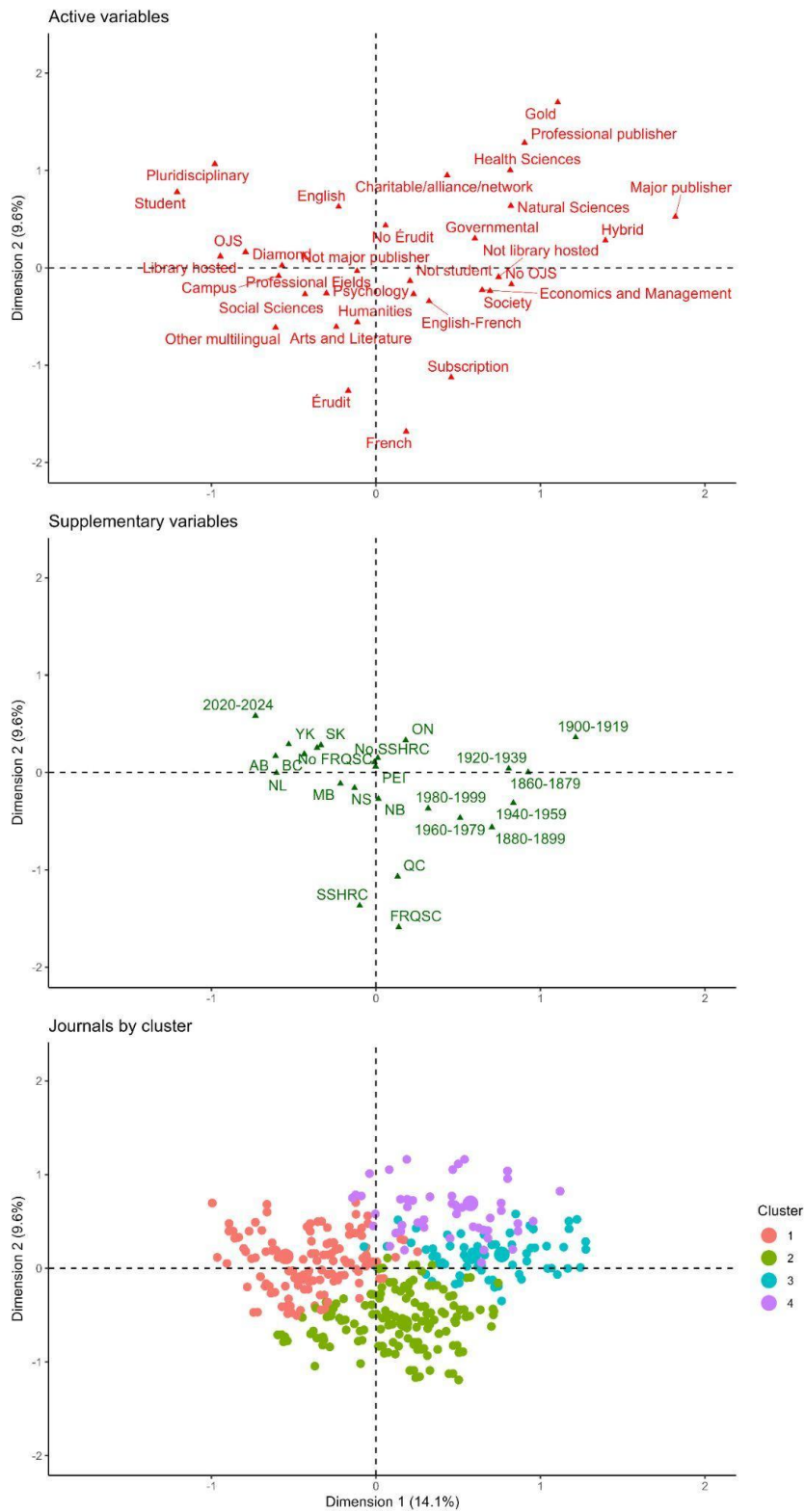

Figure 1: Canadian journal landscape according to correspondence analysis of active variables. Top panel shows linkages between active variables describing the journals. Central panel shows supplementary variables, which were excluded from the correspondence analysis, providing additional context to the pattern presented in the top panel. The bottom panel shows the individual journals projected on the same space, according to the four identified clusters, following a hierarchical clustering on the principal components resulting from the correspondence analysis.

The first cluster, which consists of 418 journals, is characterized by Diamond OA journals publishing in English. Many of these journals use OJS for management, dissemination or both. The vast majority of library-supported journals are found in this cluster, as well as student journals.

A second cluster of 257 journals typically includes French-language journals, as well as bilingual ones, using a subscription model and published by societies or on campus. These journals relatively rarely use OJS, but they are often disseminated on Érudit, with more than half being based in Quebec. Besides using subscriptions, Diamond OA is relatively frequent among these journals. They are often associated with Arts and Literature, and the Humanities. Naturally, the vast majority of journals receiving support from FRQSC is present in cluster 2, yet this is also the case for a majority of SSHRC-supported journals.

The third cluster, representing 163 journals, is mainly composed of hybrid, bilingual journals edited by learned societies, mostly active in the Natural Sciences. Journals of this group were generally the longest running, with a median year of creation of 1974. This cluster includes the vast majority of journals associated with the six major publishers mentioned previously, but also the majority of journals published by University of Toronto Press and Canadian Science Publishing.

Journals published by professional publishers are gathered in the fourth cluster. The 106 journals identified are almost exclusively in English and are associated with the Health Sciences. Being relatively young, they use Gold OA, with some journals publishing in Diamond OA. The main publisher in this cluster is JMIR Publications.

Table 3: Typology of active Canadian scholarly journals based on cluster analysis. Identified variables are meant to draw a general image of each cluster; clusters may show overlap according to some of the variables.

| Cluster number | Journals (n) | Dominant access type | Dominant language(s) | Dominant organization | OJS usage (%) | Dominant discipline | Year of creation (median) |
|---|---|---|---|---|---|---|---|
| 1 | 418 | Diamond | English | Campus | 94 | Social Sciences | 2011 |
| 2 | 257 | Subscription | English-French | Campus | 19 | Social Sciences | 1989 |
| 3 | 163 | Hybrid | English-French | Society | 6 | Natural Sciences | 1974 |
| 4 | 106 | Gold | English | Professional Publisher | 26 | Health Sciences | 2013 |

## Language

Today, the vast majority of journals either accept only English-language submissions, or English and French, especially in the Health and Natural Sciences (Figure 2). English-French bilingual journals[4], which are often managed by societies, used to dominate until the early 2000s. The growth of on-campus journals, which relatively often only allow submissions in English, has contributed to the current dominance of this language. Nevertheless, the growth of English-language journals appears to have somewhat stagnated over the last decade in SSH disciplines.

French-language journals have been more common in the Social Sciences and Humanities and associated disciplines, at least since 1940, compared to the Health and Natural Sciences. The share of these journals has remained relatively stable at around 10%. In the Health and Natural Sciences, French-language journals have become more rare since the 1950s, currently accounting for 3% of the journals. Plurilingual journals, defined as allowing languages other than English and French, show a slow but steady increase, today representing 5% of the journals in the Social Sciences and Humanities disciplines; yet only one such journal was found in the Health and Natural Sciences.

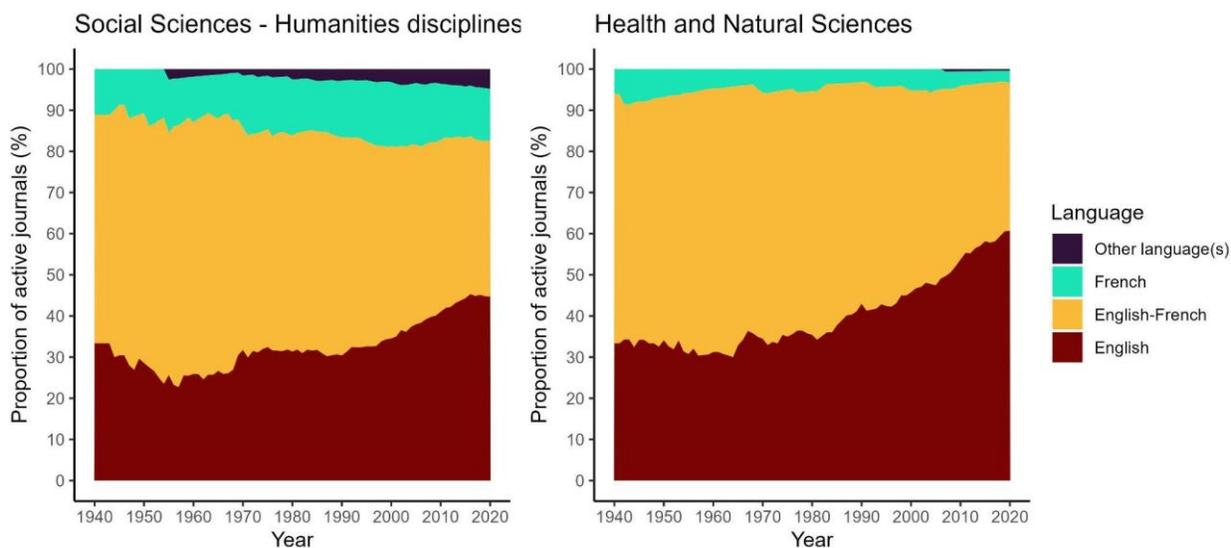

Figure 2: Proportions of active journals per year and per language in Social Sciences and Humanities disciplines (left panel) and in Health and Natural Sciences (right panel).

## Access models

The vast majority of journals active today have adopted OA, with Diamond OA by far the most common model, used by 62% of the journals. Gold OA is being used by 7%, while hybrid journals, which may provide access to a subset of their collection, represent 14% of the active periodicals. We found that 16% of the journals still demand a subscription to access parts or the entirety of their collection, yet some may allow for Green OA, i.e. they allow authors to make a version of the paper accessible through a repository.

---

[4] The fact that bi- or multilingual journals accept submissions and publish articles in different languages does not necessarily imply a balanced distribution of those languages. The figures for language distribution may be different if the analyses were conducted at the level of the article.

Recently created Canadian journals have a strong penchant for Diamond OA, with Gold OA as a second option, as out of the 221 journals founded since 2015 (and still active today), 187, or 85%, use the Diamond OA model (Figure 3). Gold OA represents 15%, and only one journal uses a subscription model. Only 22% of the journals that currently offer Diamond or Gold OA were founded before 1995. Assuming that the general uptake of OA publishing started in the mid-1990s, we infer that at least a quarter of current OA journals active today were not created as such, but 'flipped' to become OA during the last decades. This proportion may be higher as some of the younger OA journals, i.e. created after 1995, may have used a subscription model during their first years of activity. The vast majority of journals that currently use hybrid or subscription models were founded before 1995, at 90% and 80%, respectively (Figure 3). Based on these numbers, we may conclude that most of these journals started as subscription journals, with some switching to hybrid publishing along the way.

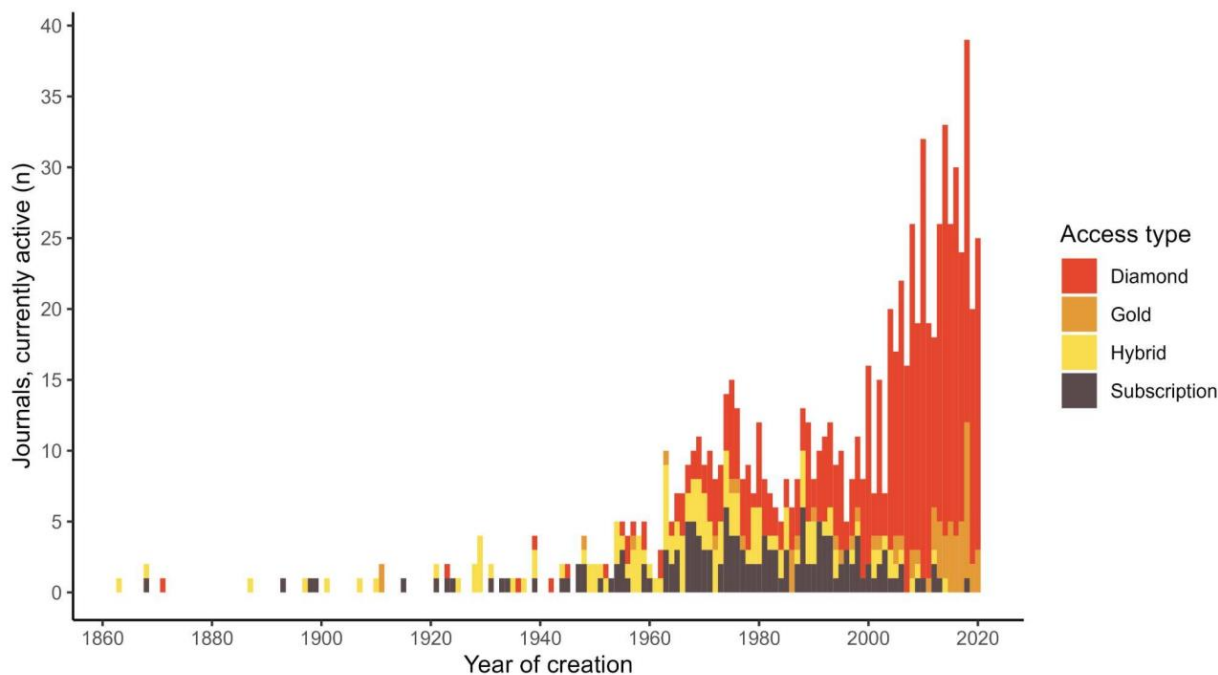

Figure 3: Access types of currently active journals, according to the year of creation.

Of the active journals theoretically admissible for inclusion in the Directory of Open Access Journals (DOAJ), i.e. active Diamond and Gold OA journals, 26% of the Diamond OA and 54% of the Gold OA journals are indeed indexed. Comparing the different organization types, OA journals of professional publishers are most frequently indexed in DOAJ, at 52% of potentially admissible journals. OA journals affiliated to universities show the lowest level of indexing in DOAJ, at 25%, and this proportion is particularly low for student OA journals, of which only 6% is indexed.

# Historical perspectives and emerging trends in journal creation and cessation

The number of active Canadian scholarly periodicals has increased from about a dozen in 1900 to more than 900 today (Figure 4). Whereas growth was limited until the 1950s, when close to a hundred journals were being published nationwide, many journals were founded during the 1960s and 1970s. Before this period the majority of journals were associated with the Natural and Health Sciences; the newly founded journals, however, were much more often associated with the Social Sciences, Humanities, Professional Fields and Arts and Literature.

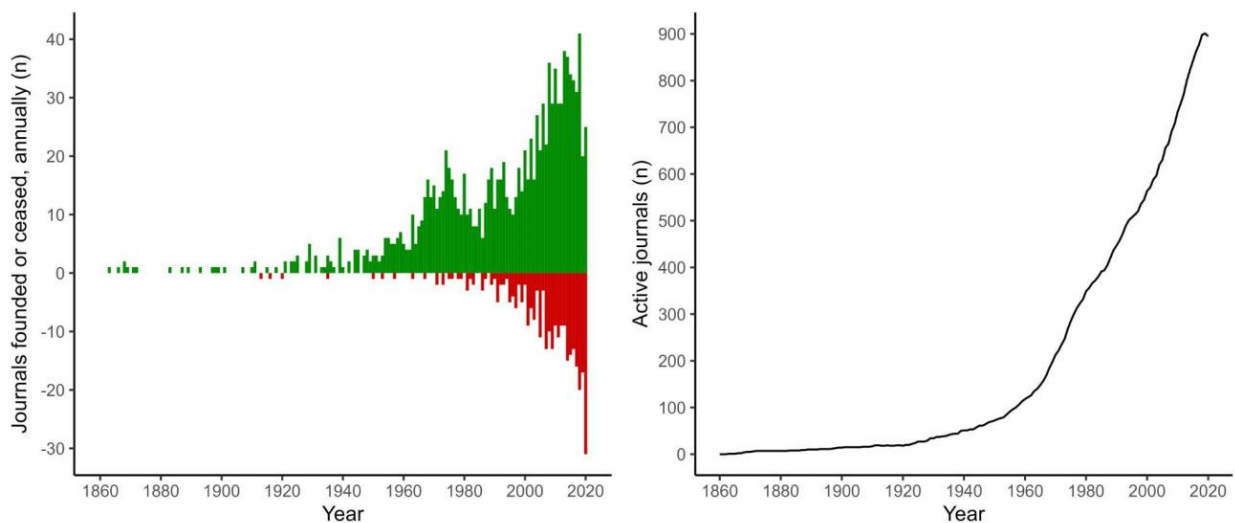

Figure 4: Evolution of the Canadian journal landscape since the mid-19th century. Left panel shows the annual number of journals founded (green) or ceased (red). Right panel shows the number of active journals from 1860 to 2020.

During the 1980s, the creation of new journals slowed down and remained low until the early 2000s. Likely associated with the accessibility of online publishing, the number of journals founded increased rapidly in the new millennium, with 30 to 40 new journals annually. However, the trend in journals being founded was mirrored by that of journals ceasing publication. During the most recent years, the number of ceasing journals almost equalizes the number of newly created ones, although there may be a delay effect for some very recently created journals that have not been included in our dataset.

Analyzing the entire dataset, i.e. covering more than 150 years, we found 312 periodicals having ceased publication, equivalent to 25% of the total number of journals having been active during this period. Furthermore, English-language journals appear to have a particular instability: 30% of those that were in activity at one point in time have now ceased publication, compared to 20% of French-language journals and 19% of English-French bilingual journals. Focusing on journal creation and cessation during the digital era, it appeared journals publishing in English and managed on campus have had the highest rates of both creation and cessation (Figure 5). Geographically, journal creation and cessation were greatest in Alberta and British Columbia, intermediate in Ontario and lowest in Quebec. Any effects of discipline on journal cessation were not detected, as the distribution of ceased journals did not appear to be significantly different from the distribution of active journals. Since accurate OA status data was not available for ceased journals, we could not quantify journal cessation as a function of OA status. Nevertheless, main characteristics of ceased journals included being managed on campus, publishing in English and origins concentrated in Alberta and British Columbia; all these traits point to journals from cluster 1 (Figure 1 and Table 3), which suggests that Diamond OA journals have likely been overrepresented among the ceased journals.

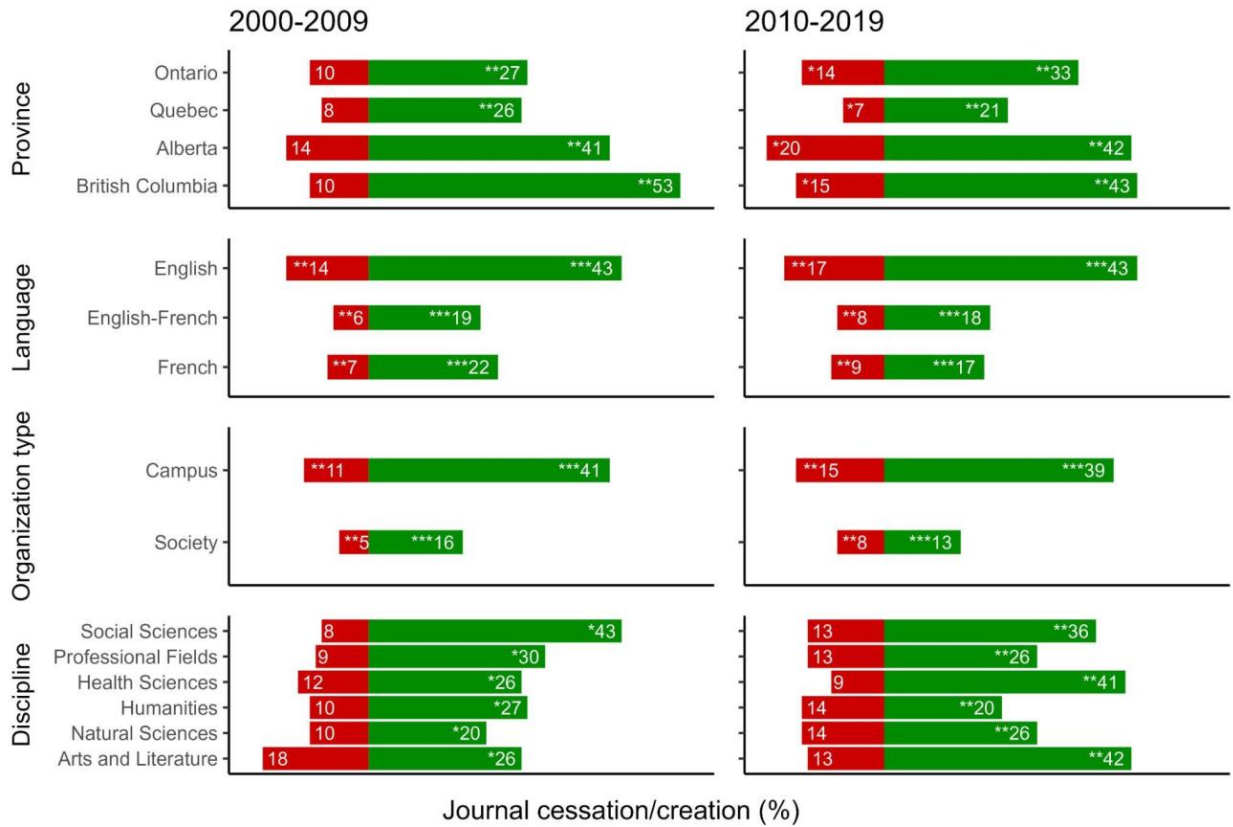

Figure 5: Prevalence of journal cessation (red) and creation (green) as a function of origin (limited to the four major provinces), language(s), organization type and discipline, for 2000-2009 and 2010-2019. Percentages show the proportion of journals having been active during the period that were created, or have ceased, during the same period. Similarities between active journal distributions and ceased/created journal distributions were determined using a Chi-squared goodness of fit test. Asterisks refer to significance levels: * at p<0.05; ** at p<0.01; *** at p<0.001.

Interestingly, the median age of active journals has hardly changed since the mid 20th century. In 1950, the median journal age was 19 years, compared to 20 years in 2020 (Figure 6). However, the digital era has been particular in showing a divergence in journal sustainability and lifespan. While the 25% youngest journals have become progressively younger since the early 2000s, the 25% oldest journals have become older. These opposing trends show that, while older journals manage to consolidate their activity, younger journals are in a cycle of ongoing renewal, characterized by high cessation rates. This is confirmed by the median age of journals upon cessation for the 2000-2020 period, which was only 11 years.

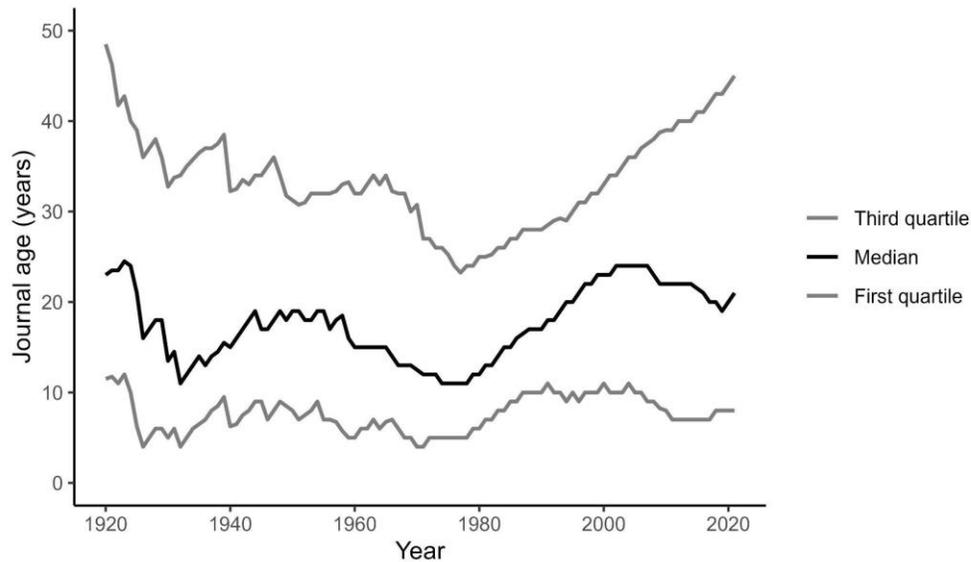

Figure 6: Evolution of journal age for active journals, as quantified by the median and first and third quartiles.

Despite changing creation and cessation dynamics since the expansion of online publishing, the relative presence of the main types of organizations (i.e. campus, society and professional publishers) pursues a trend that was initiated more than fifty years ago. Universities have become the dominant entities managing journals, while societies today account for 35% of the active journals. This trend coincides with the previously mentioned increasing presence of journals of the Social Sciences, Humanities, Professional Fields and Arts and Literature.

Professional publishers have also increased their share, especially during the last decade, currently representing around 9% of the journal publishers. The relative number of journals affiliated to government-related organizations has decreased during the digital era, accounting for just over 1% of the total (Figure 7).

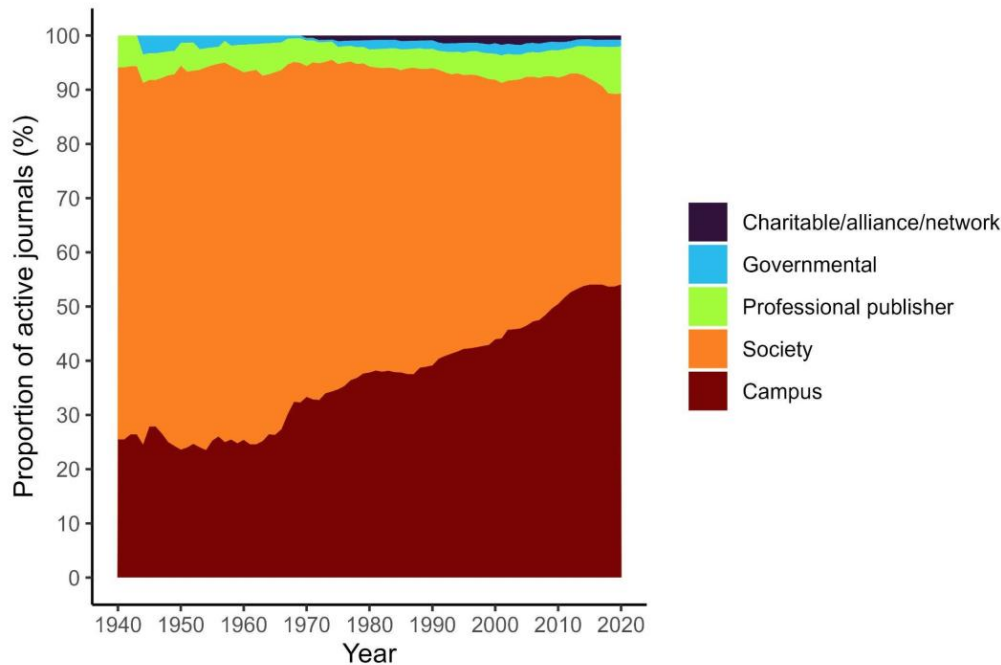

Figure 7: Proportion of active journals per year and per type of organization.

## Discussion

The results of this study are consistent with previous studies which have attempted to characterize the Canadian publishing landscape. Its main traits are the massive adoption of Diamond OA, the predominance of the Social Sciences and Humanities disciplines, and the sparse presence of the major commercial publishers. Canadian scholarly journals also present a great diversity in terms of the role of supporting organizations and publishers, language, and journal lifespan, which diverge along geographic gradients (Figure 1). Patterns related to journal origin can partly be explained by their linguistic context, but they may also reflect historical and cultural differences in the organization of scholarly publishing and the roles of the actors involved. Library support for journal publishing, for example, has become a standard practice for many Canadian universities (Whyte Appleby et al., 2018). Since equitable access to information is one of the core values of libraries, especially with regards to OA publishing and the persisting repercussions of the 'serials crisis', it is not surprising librarians have been particularly active in supporting journals (Lippincott, 2017). Today, libraries are heavily involved in journal support in all provinces, particularly important in Alberta and British Columbia, but much less so in Quebec. The development of library support for journals may have reinforced the use of OJS, and vice versa (Richard et al., 2009). Meanwhile, the more limited development of library publishing in Quebec may be linked to the creation, in 1998, of Érudit, and its mission of supporting journals, with almost 250 active journals disseminated at the end of 2024. It may also be associated with the particular role of librarians in Quebec, who, unlike their colleagues from the other provinces, often have a professional status rather than an academic one. As a result, Quebec librarians' roles may be described as being 'supporting' and 'administrative' and they are rarely actively involved in research (Zavala Mora et al., 2022). These librarians may thus be less connected to the publishing process in general and editorial support in particular. Finally, the financial and infrastructural support of FRQSC (and its ancestors) for Quebec journals may also explain why direct library support to journals has not been developed as much as in the rest of Canada.

## Open access mandates and uptake

Both the current SSHRC Aid to Scholarly Journals program and the FRQSC *Soutien aux revues scientifiques* aim to support SSH journals that are entirely OA (thus excluding hybrid journals) or that have a 12-month maximum embargo to access. These journals sell subscriptions for users to access the most recent content. In practice, both SSHRC and FRQSC programs particularly support journals using the latter model. Our data show that SSHRC provides funding for 30% of the Canadian subscription journals in SSH and 12% of Diamond OA journals of the same disciplines. In Quebec, FRQSC supports 48% of subscription journals (publishing in French, or French and English), compared to 40% of Diamond OA journals. However, many subscription journals require a subscription for their entire collection, or they have embargo periods exceeding 12 months. This makes them ineligible for funding, which means the denominator in our calculation may be overestimated; by consequence the success rates for subscription journals to obtain support are underestimated. In any case, these numbers show that a subscription journal currently stands a better chance of being funded than a Diamond OA journal.

Like FRQSC, the SSHRC program particularly supports journals from Quebec, in comparison with the three other main provinces: 37% of Quebec SSH journals are effectively supported, compared to 9% for Ontario, 8% for British Columbia and 4% for Alberta, despite British Columbia and Alberta having much higher proportions of Diamond OA journals, both in general (Figure 3), but also specifically in SSH. It is not possible to identify any bias in the attribution of funding, because no (public) data on journals applying for SSHRC or FRQSC funding is available. Nevertheless, the current programs may fall short in their respective aims to encourage and promote the transition to OA models of publishing. As subscription models, with or without embargo, remain frequent in Quebec, in comparison with the other provinces (Figure 3), we hypothesize that the past and current support programs may have contributed to journals adopting embargos or consolidating them, rather than flipping to OA.

## Journal sustainability

We found that 25% of the Canadian scholarly journals active since the mid-19th century have ceased publication. This proportion is consistent with a 2017 report by the Canadian Scholarly Publishing Working Group, which estimated that around 20% of Canadian journals ever founded had ceased publication by that year (Canadian Scholarly Publishing Working Group, 2017). The widespread adoption of online publishing has allowed for a greater potential for dissemination along with reduced expenses, and it has led to new journal management tools and publishing modes as reflected by various types of OA being adopted today. However, achieving financial sustainability remains a challenge for many journals, particularly Diamond OA ones. While some journals may succeed in resorting to a wide range of funding sources, they most often rely on volunteer work, in-kind support from research performing organizations (mainly universities) and funds from government agencies (Bosman et al., 2021). Additional observations may be made for the different journal clusters we identified (Table 3); we will focus on the two major clusters, as these are likely to represent the more vulnerable journals.

### Cluster 1 journal sustainability

The journals from cluster 1 are characterized by the dominance of English, the frequent use of OJS, the links with universities and a general presence of library support, the adoption of Diamond OA and origins in Alberta, British Columbia and, to a lesser extent, Ontario. Among the four clusters, these journals have shown both the highest rates of creation and cessation since the digital era (Figure 5), and cessation prevalence is particularly high for the younger journals (Figure 6). When comparing internationally, the cessation rates of Canadian independent OA journals may still be relatively low; focusing on 250 scholar-published, OA journals founded before 2002, Björk et al. (2016) found only just over half of these were still active in 2014.

The availability and ease of use of digital tools, and particularly OJS, which has lowered the threshold for the creation of new journals (Björk et al., 2016), may well have had a downside concerning longevity, as many of these new journals may have been founded without much planning or perspectives for long-term support. This may partly explain the short life cycle of the journals in cluster 1. Language may also be playing a role in this tendency. English-language journals have ceased publication much more frequently than journals allowing French or other languages (Figure 5). In a context of the ever-growing presence of English in research outputs, this may be counterintuitive - one would rather expect the number of journals publishing in French to be declining. However, linguistic market effects may explain this tendency. Canadian non-profit English-language journals operate in an international market, where journals compete in highly unequal conditions to attract authors, accrue scientific and symbolic capital, and ensure financial resources for their stability. Unlike their French-language counterparts, these journals cannot exploit their linguistic particularity to effectively attain a national readership. On the other hand, French-language Canadian journals may be able to capitalize on their national focus through language, allowing researchers to attain a readership that may be best addressed in French, particularly presenting research topics with a more specific cultural or geographic relevance (Larivière, 2018; Van Bellen & Larivière, 2024). It should be noted, however, that the proportion of English-language journals continues to increase slightly (Figure 2).

### Cluster 2 journal sustainability

Cluster 2 is characterized by scholarly journals using subscriptions, or alternatively Diamond OA, and they are frequently disseminated on Érudit. Compared to cluster 1, these journals receive more frequent support from SSHRC and FRQSC, or both, and they have generally been running for a much longer period. With a few exceptions, all French-language journals are gathered in this cluster.

The journals from cluster 2 have shown very low cessation rates during the last decades; in the meantime, few journals have been created. Compared with cluster 1 journals, the journals of cluster 2 are characterized by different types of support. Both government programs, which have been in effect for more than 40 years, are likely to have had a stabilizing effect on SSH journals.

All journals receiving FRQSC support, which has both a financial and an infrastructural component, are disseminated on Érudit. Besides ensuring dissemination, Érudit actively supports journals through production services, documenting and registering metadata, optimizing discoverability and general advice on standards and best practices in scholarly publishing. Access to publishing and archiving infrastructures and technical and financial support may well be key factors in sustaining a digital journal's trajectory over time. However, there may be a kind of Matthew effect in place, whereby more established journals with experienced editorial boards are in a better position to successfully apply for funding, and, if funded by FRQSC, become integrated on Érudit. The policies and programs of the province of Quebec and other initiatives related to the revitalization of French as a language of science may also be protecting this linguistic and academic niche from the centripetal forces that pressure scholars towards publishing "in high-profile centre journals instead of in their traditional local or regional outlets" (Bennett, 2014, p. 241).

## Future research and policy perspectives

In the light of the non-negligible figures of inactive journals found in this work, more research is needed to better understand the factors that determine both the creation and the cessation of journal activities. The total number of active journals may still be on the rise (as shown by Figure 4) and journals ceasing activity is not a new phenomenon (Figure 6), but journal cessation may still have multiple downsides, such as a loss of editorial expertise and resources, and it may lead to vanishing scholarly content, especially for electronic journals lacking a long-term digital preservation policy (Laakso et al., 2021). We call for a deeper understanding of the needs of journals with different profiles in order to provide them with more tailored support and thus ensure their economic sustainability and technical preservation in time.

In December 2023, Canada's three main federal research granting agencies (CIHR, NSERC and SSHRC) announced they would undertake a review of their OA policy for researchers. Once in effect, by the end of 2025, the *Tri-Agency Open Access (OA) Policy on Publications* will require that "any peer-reviewed journal publications arising from agency-supported research be freely available, without subscription or fee, at the time of publication" (Adem et al., 2023). Earlier, in 2021, the FRQ joined cOAlition S, and globally aligned its policy with that of Plan S in 2022. Compared to its previous 2019 policy, it presents three main changes: "[supported researchers'] articles must be made OA immediately (rather than within a twelve-month timeframe), they must bear one of the two more open Creative Commons licences (CC-BY, CC-BY-ND, or equivalent), and hybrid OA articles fees are no longer grant-eligible expenses except under certain conditions" (Harris et al., 2024).

Currently, both SSHRC and FRQ are adapting their journal support programs, likely in line with the revised researcher mandates. Requiring journals to adopt OA while excluding subscription (and hybrid) models should favour Diamond OA. Recently, FRQ have announced that supported journals will be required to apply to be indexed in DOAJ. To what extent the reviewed mandates will lead to changes in the group of supported journals remains difficult to predict, as, besides the compliance to requirements, both the urgency of obtaining financial support and the readiness to apply will influence the ultimate pool of candidates. It will also depend on the ease of (currently supported) subscription journals to flip to a compliant OA model. As noted earlier, our analyses show that almost a quarter of active OA journals have 'flipped' from a subscription model, in line with a previous estimate specifically for library-hosted journals of 16% (Willinsky, 2017); these numbers suggest this is not a marginal pathway to OA. The effects of the new OA policies on researchers' publishing practices and journals' economic planning would need a critical assessment in the future, also because compliance to researcher mandates has been lower in Canada than the United States and many European countries (Robinson-Garcia et al., 2020; Simard et al., 2022).

### Limitations

This study encountered a few limitations. The first one was related to the identification of changes through time. Even though it was relatively straightforward to identify the year a journal was founded, it was highly complex, or impossible, to identify changes in journal characteristics throughout its history. We used Library and Archives Canada's Aurora catalog as a main source for information on journals, but its records do not allow to recover any changes in the publishing organization or language policies. Any detailed historic information had to be recovered from journal websites, but in many cases details were difficult to find or absent. Likewise, due to time and resource constraints, it was not possible to establish the years journals joined commercial publishers, or the years journals switched access modes.

The definition and the identification of the origin of a journal is debatable and may change throughout its history. We encountered a few journals which used to be published by a Canadian society, before joining an international partner organization and losing their Canadian scope. Such journals were omitted from the study.

The interest in documenting patterns in Canadian scholarly journal characteristics lies partly in the role of these journals for national dissemination of knowledge. Naturally, some journals in our dataset play a much greater role in this respect than others and a few journals are likely weakly embedded in Canadian research networks. Accurately defining the degree of "nationality" of a journal ideally requires analyzing author- and/or article-level data. These types of analyses were not feasible in the context of this work, yet the availability of a national journal dataset is a requirement for exploring article metadata in detail, which opens the door to future research.

## Conclusion

The different journal profiles identified in this study speak of a highly diversified landscape across Canadian provinces, OA types, supporting organizations and languages. These journals support bibliodiversity in terms of the diversity of the authors who publish in them, the research subjects they present, their publishing organizations and the languages used.

Journal publishing practices have evolved over time, with clear transformations over the last decades since the advent of digitalization. In light of our data, we argue that online publishing has markedly changed journal management, production and dissemination, aided by open tools and platforms. In Canada, the development of OJS and the creation of Érudit, which are different approaches to supporting journals, have proven effective in maintaining a diversified national publishing landscape. Diamond OA has been adopted by 62% of the active journals, and the vast majority of the journals founded during the last decade are using this model today.

Nevertheless, journal cessation has been a growing issue since the start of the digital era, and this disproportionately affects younger journals. Resources for starting journals may be inadequate, especially regarding the stability of staff and financial support. However, as many journals cease publication within a few years following creation, we also suspect a more critical evaluation of the minimal requirements for founding a journal would lead to a higher proportion of successful launches. To this effect, starting journal editors would require a long-term perspective for support, allowing them to enhance their level of indexing to ensure discoverability, to become compliant with the SSHRC and FRQ funding programs and to assist in the DOAJ application process.

The current beneficiaries from government funding programs, whether at federal or provincial level, have a very specific profile, represented by the cluster 2 journals in our analysis. This study provides insights for the design of more tailored policies that cater to the needs of under-resourced periodical types, and that take account of evolving practices among the entirety of the Canadian scholarly journals.

## Acknowledgements

The authors wish to thank Suzanne Beth and Jeanette Hatherill for their valuable comments and proofreading of the early versions of this manuscript.

## Conflict of interest

## References

Adem, A., Hewitt, T., & Strong, M. (2023, July 4). *The presidents of Canada's federal research granting agencies announce a review of the Tri-Agency Open Access Policy on Publications*. https://science.gc.ca/site/science/en/interagency-research-funding/policies-and-guidelines/open-access/presidents-canadas-federal-research-granting-agencies-announce-review-tri-agency-open-access-policy

Basson, I., Simard, M.-A., Ouangré, Z. A., Sugimoto, C. R., & Larivière, V. (2022). The effect of data sources on the measurement of open access: A comparison of Dimensions and the Web of Science. *PLOS ONE*, *17*(3), e0265545. https://doi.org/10.1371/journal.pone.0265545

Bennett, K. (2014). Conclusion: Combating the Centripetal Pull in Academic Writing. In K. Bennett (Ed.), *The Semiperiphery of Academic Writing: Discourses, Communities and Practices* (pp. 240–246). Palgrave Macmillan UK.

https://doi.org/10.1057/9781137351197_14

Björk, B.-C., Shen, C., & Laakso, M. (2016). A longitudinal study of independent scholar-published open access journals. *PeerJ*, *4*, e1990. https://doi.org/10.7717/peerj.1990

Boismenu, G., & Beaudry, G. (2002). *Le nouveau monde numérique: Le cas des revues universitaires*. Presses de l'Université de Montréal. https://doi.org/10.4000/books.pum.9258

Bosman, J., Frantsvåg, J. E., Kramer, B., Langlais, P.-C., & Proudman, V. (2021). The OA diamond journals study. Part 1: Findings. *Zenodo*. https://doi.org/10.5281/zenodo.4558704

Canadian Scholarly Publishing Working Group. (2017). *Canadian Scholarly Publishing Working Group Final Report*. https://www.carl-abrc.ca/wp-content/uploads/2017/07/CSPWG_final_report_EN.pdf

Céspedes, L. (2021). Latin American journals and hegemonic languages for academic publishing in Scopus and Web of Science. *Trabalhos Em Linguística Aplicada*, *60*. https://doi.org/10.1590/010318138901311520201214

Fortin, A. (2018). Penser au Québec, penser le Québec. De quelques revues de sciences sociales. *Recherches sociographiques*, *59*(3), 411–433. Érudit. https://doi.org/10.7202/1058721ar

Fyfe, A., Coate, K., Curry, S., Lawson, S., Moxham, N., & Røstvik, C. M. (2017). *Untangling academic publishing: A history of the relationship between commercial interests, academic prestige and the circulation of research*. https://doi.org/10.5281/zenodo.546100

Gingras, Y. (2014). *Les dérives de l'évaluation de la recherche: Du bon usage de la bibliométrie*. Raisons d'agir.

Gingras, Y., & Mosbah-Natanson, S. (2010). Les sciences sociales françaises entre ancrage local et visibilité internationale. *European Journal of Sociology / Archives Européennes de Sociologie*, *51*(2), 305–321. Cambridge Core. https://doi.org/10.1017/S0003975610000147

Gordon, W. (1984). *A Study of the Canadian Periodical Publishing Industry: A Report*. Department of Communications.

Harris, R., Lange, J., & Lasou, P. (2024). Plan S and Open Access (OA) in Quebec: What Does the Revised FRQ OA Policy Mean for Researchers? *Evidence Based Library and Information Practice*, *19*(1), 35–57. https://doi.org/10.18438/eblip30413

Krapež, K. (2023). Impact of publisher's commercial or non-profit orientation on editorial practices: Moving towards a more strategic approach to supporting editorial staff. *Learned Publishing*, *36*(4), 543–553. https://doi.org/10.1002/leap.1575

Laakso, M., Matthias, L., & Jahn, N. (2021). Open is not forever: A study of vanished open access journals. *Journal of the Association for Information Science and Technology*, *72*(9), 1099–1112. https://doi.org/10.1002/asi.24460

Lange, J., & Severson, S. (2019). *List of Canadian independent, scholarly journals (2020)* [Dataset]. Borealis. https://doi.org/10.7939/DVN/EPSJJR

Lange, J., & Severson, S. (2021). What Are the Characteristics of Canadian Independent, Scholarly Journals? Results from a Website Analysis. *Journal of Electronic Publishing*, *24*(1). https://doi.org/10.3998/jep.153

Lange, J., & Severson, S. (2022). Work It: Looking at Labour and Compensation in Canadian Non-Commercial Scholarly Journals. *KULA: Knowledge Creation, Dissemination, and Preservation Studies*, *6*(2), 1–21. https://doi.org/10.18357/kula.151

Larivière, V. (2018). Le français, langue seconde? De l'évolution des lieux et langues de publication des chercheurs au Québec, en France et en Allemagne. *Recherches sociographiques*, *59*(3), 339–363. Érudit. https://doi.org/10.7202/1058718ar

Larivière, V., Beth, S., van Bellen, S., Delmas, È., & Paquin, É. (2021). *Les revues savantes canadiennes en sciences humaines et sociales—Portrait quantitatif et qualitatif* (p. 47). https://www.erudit.org/public/documents/Revues_canadiennes_shs_2021.pdf

Larivière, V., Haustein, S., & Mongeon, P. (2015). The Oligopoly of Academic Publishers in the Digital Era. *PLOS ONE*, *10*(6), e0127502. https://doi.org/10.1371/journal.pone.0127502

Larivière, V., & Warren, J.-P. (2019). The Dissemination of National Knowledge in an Internationalized Scientific Community. *The Canadian Journal of Sociology/Cahiers Canadiens de Sociologie*, *44*(1), 1–8.

Lippincott, S. K. (2017). *Library As Publisher: New Models of Scholarly Communication for a New Era* (Against the Grain (Media), LLC). https://doi.org/10.3998/mpub.9944345

Lorimer, R., & Lindsay, A. (2004). Canadian scholarly journals at a technological crossroads. *Canadian Journal of Communication*, *29*(3), 253–276. https://doi.org/10.22230/cjc.2004v29n3a1467

Ma, Z. (2019). The Relevance of National Journals from a Chinese Perspective. In W. Glänzel, H. F. Moed, U. Schmoch, & M. Thelwall (Eds.), *Springer Handbook of Science and Technology Indicators* (pp. 505–562). Springer International Publishing. https://doi.org/10.1007/978-3-030-02511-3_20

Moed, H. F., de Moya-Anegon, F., Guerrero-Bote, V., Lopez-Illescas, C., & Hladchenko, M. (2021). Bibliometric assessment of national scientific journals. *Scientometrics*, *126*(4), 3641–3666. https://doi.org/10.1007/s11192-021-03883-5

Morrison, H. (2016). Small scholar-led scholarly journals: Can they survive and thrive in an open access future? *Learned Publishing*, *29*(2), 83–88. https://doi.org/10.1002/leap.1015

Paquin, É. (2016). *Shaping a Collective Future: An Investigation into Canadian Scholarly Journals' Socio-Economic Reality and an Outlook on the Partnership Model for Open Access* (p. 32). Université de Montréal. https://www.erudit.org/public/documents/Canadian_Journals_Socio-Economic_Study.pdf

Piwowar, H., Priem, J., Larivière, V., Alperin, J. P., Matthias, L., Norlander, B., Farley, A., West, J., & Haustein, S. (2018). The state of OA: a large-scale analysis of the prevalence and impact of Open Access articles. *PeerJ*, *6*, e4375. https://doi.org/10.7717/peerj.4375

Pölönen, J., Syrjämäki, S., Nygård, A.-J., & Hammarfelt, B. (2021). Who are the users of national open access journals? The case of the Finnish Journal.fi platform. *Learned Publishing*, *34*(4), 585–592. https://doi.org/10.1002/leap.1405

Priem, J., Piwowar, H., & Orr, R. (2022). *OpenAlex: A fully-open index of scholarly works, authors, venues, institutions, and concepts* [Computer software]. https://arxiv.org/abs/2205.01833

Richard, J., Koufogiannakis, D., & Ryan, P. (2009). Librarians and libraries supporting open access publishing. *Canadian Journal of Higher Education*, *39*(3), 33–48. https://doi.org/10.7939/R36S2D

Robinson-Garcia, N., Costas, R., & van Leeuwen, T. N. (2020). Open Access uptake by universities worldwide. *PeerJ*, *8*, e9410. https://doi.org/10.7717/peerj.9410

Simard, M.-A., Butler, L.-A., Alperin, J. P., & Haustein, S. (2024). We need to rethink the way we identify diamond open access journals in quantitative science studies. *Quantitative Science Studies*, 1–5. https://doi.org/10.1162/qss_c_00331

Simard, M.-A., Ghiasi, G., Mongeon, P., & Larivière, V. (2022). National differences in dissemination and use of open access literature. *PLOS ONE*, *17*(8), e0272730. https://doi.org/10.1371/journal.pone.0272730

Sivertsen, G. (2016). Patterns of internationalization and criteria for research assessment in the social sciences and humanities. *Scientometrics*, *107*(2), 357–368. https://doi.org/10.1007/s11192-016-1845-1

Van Bellen, S. (2024). *Revues savantes canadiennes/Canadian scholarly journals* (Version V1) [Dataset]. Borealis. https://doi.org/10.5683/SP3/9ONCEU

Van Bellen, S., Alperin, J. P., & Larivière, V. (2024). The oligopoly of academic publishers persists in exclusive database. *arXiv Preprint arXiv:2406.17893*.

Van Bellen, S., & Larivière, V. (2024). Les revues canadiennes en sciences sociales et humaines: Entre diffusion nationale et internationalisation. *Recherches sociographiques*, *65*(1), 15–35. Érudit. https://doi.org/10.7202/1113756ar

Whyte Appleby, J., Hatherill, J., Kosavic, A., & Meijer-Kline, K. (2018). What's in a Name? Exploring identity in the field of library journal publishing. *Journal of Librarianship and Scholarly Communication*, *6*(1). https://doi.org/10.7710/2162-3309.2209

Willinsky, J. (2017). Modelling a Cooperative Approach to Open Access Scholarly Publishing: A Demonstration in the Canadian Context. *Canadian Journal of Communication*, *42*(5), 923–934. Canadian Business & Current Affairs Database; Sociological Abstracts. https://doi.org/10.22230/cjc.2017v42n5a3264

Zavala Mora, D., Quirion, G., Biondo, S., & Bouchard, J. (2022). Le statut universitaire pour les bibliothécaires de l'Université Laval: Une question d'équité entre les bibliothécaires universitaires canadiens. *Canadian Journal of Academic Librarianship / Revue canadienne de bibliothéconomie universitaire*, *8*, 1–37. Érudit. https://doi.org/10.33137/cjal-rcbu.v8.38458